\documentclass[sigconf]{acmart}

\usepackage{graphicx} 
\usepackage{mathtools} 
\usepackage{float}
\usepackage[ruled,vlined,linesnumbered]{algorithm2e}
\usepackage{enumitem}
\newlist{Properties}{enumerate}{2}
\setlist[Properties]{label=Property \arabic*.,font=\textbf,itemindent=*}

\usepackage{float}

\theoremstyle{definition}

\allowdisplaybreaks
\newcommand{\ie}{{\it i.e.} }

\newcommand{\eg}{{\it e.g., }}
\newcommand{\xeg}{{\it E.g.,}}

\newif\iflong

\title{On Recommending Category: A Cascading Approach}

\author{Qihao Wang}
\authornote{Equal contribution.}
\affiliation{%
  \institution{University of Illinois at Urbana-Champaign}
  \country{Urbana, IL, USA}
}
\email{qwang65@illinois.edu}

\author{Pritom Saha Akash}
\authornotemark[1]
\affiliation{%
  \institution{University of Illinois at Urbana-Champaign}
  \country{Urbana, IL, USA}
}
\email{pakash2@illinois.edu}

\author{Varvara Kollia}
\affiliation{%
  \institution{eBay Inc.}
  \country{San Jose, CA, USA}
}
\email{rkollia@ebay.com}

\author{Kevin Chen-Chuan Chang}
\affiliation{%
  \institution{University of Illinois at Urbana-Champaign}
  \country{Urbana, IL, USA}
}
\email{kcchang@illinois.edu}

\author{Biwei Jiang}
\affiliation{%
  \institution{eBay Inc.}
  \country{San Jose, CA, USA}
}
\email{bjiang1@ebay.com}

\author{Vadim Von Brzeski}
\affiliation{%
  \institution{eBay Inc.}
  \country{San Jose, CA, USA}
}
\email{vbrzeski@ebay.com}

\begin{document}
\longtrue

\begin{abstract}
Recommendation plays a key role in e-commerce, enhancing user experience and boosting commercial success. Existing works mainly focus on recommending a set of items, but online e-commerce platforms have recently begun to pay attention to exploring users' potential interests at the category level. Category-level recommendation allows e-commerce platforms to promote users' engagements by expanding their interests to different types of items. In addition, it complements item-level recommendations when the latter becomes extremely challenging for users with little-known information and past interactions. Furthermore, it facilitates item-level recommendations in existing works. The predicted category, which is called intention in those works, aids the exploration of item-level preference. However, such category-level preference prediction has mostly been accomplished through applying item-level models. Some key differences between item-level recommendations and category-level recommendations are ignored in such a simplistic adaptation. In this paper, we propose a cascading category recommender (CCRec) model with a variational autoencoder (VAE) to encode item-level information to perform category-level recommendations. Experiments show the advantages of this model over methods designed for item-level recommendations.

\end{abstract}

\maketitle

\section{Introduction}
\label{sec:intro}
\begin{figure}[h]
\centering
\includegraphics[width=\linewidth]{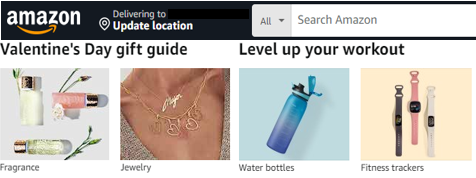}
\caption{A Snapshot of Category Recommender}
\label{fig:real_amazon}
\end{figure}

Recommender \cite{attention,fmlp,sasrec,liang2018variational,wu2016collaborative,he2017neural,tay2018latent,wang2019neural,sankar2021protocf,park2012adaptive,yin2012challenging} systems from existing works mainly focus on recommending a particular set of items, but there has recently been a growing interest in category-level recommendation. In category-level recommendation, instead of directly recommending products, users are recommended the categories of products. Major platforms like Amazon and eBay already feature category recommendations separately from item recommendations. For example, Figure.~\ref{fig:real_amazon} shows a snapshot from Amazon, where categories are recommended directly to the users. 


\begin{figure}[h]
\centering
\includegraphics[width=\linewidth]{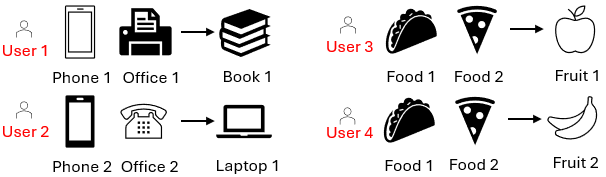}
\caption{User Interactions in an E-commerce Website}
\label{fig:user_example}
\end{figure}

Traditionally, category recommendation has been explored only to aid item recommendation \cite{air,intention1,intention2,intention3}. These works leverage \textit{category-level preferences}, or "intentions", to improve item-level recommendations, especially for new users facing the cold-start problem. Category-level preferences are generally more stable and less varied than item-level interests, as users often explore within a narrower range of categories despite showing interest in a broad array of items. This stability makes category-level signals more reliable, reducing the likelihood of overlooking preferred categories, unlike the more volatile item-level interactions.

Moreover, category recommendation is important in its own right, guiding users' high-level interaction in e-commerce. 
In particular, it offers \textit{exploration and navigation of e-commerce systems}: An online store may highlight categories of interest at the top of its website to lead users to navigate into different departments, as Fig. ~\ref{fig:real_amazon} shows. Other benefits include enhanced \textit{fairness of recommendations} and support for \textit{platform campaigns}. We discuss the motivations for category recommendation in Section 3.

Despite the increasing demands to explore categorical interests, category-level recommendations remain relatively under-explored within the literature. Previous studies \cite{air,intention2,intention3} predominantly adapt methodologies designed for item-level recommendations to category-level prediction. \xeg the model proposed in \cite{air} leverages an attention-based framework closely mirroring those employed for item-level recommendation \cite{attention}. Similarly, \cite{intention3} feeds categories directly to the Temporal Convolutional Network in place of items.

\noindent \textbf{Problem.} This paper introduces the problem of \textit{category-level recommendation} (CatRec), highlighting two key distinctions from traditional item-level recommendation systems. First, \textbf{category-level interactions are indirect}. Unlike item-level recommendations, where users interact directly with specific items, CatRec requires inferring a user's interest in broader categories based on their interactions with individual items. For example, if a user interacts with items like Phone1, Phone2, and Book1, CatRec would infer interests in the categories "Phones" and "Books," as illustrated in Fig. ~\ref{fig:user_example}. Second, \textbf{the space for CatRec is constrained}. As CatRec aids in exploring and navigating e-commerce platforms, they are mainly used at the top of e-commerce homepages or emails (\eg Fig.~\ref{fig:real_amazon}), where only limited spaces are available.



\noindent \textbf{Challenges.} Due to the above distinctive phenomena, CatRec introduces the following three challenges.\\
\textbf{Indistinguishable Category Interaction.} Inferred category interactions can be limiting in distinguishing between users, as they generalize interactions using broader categories. Even if two users have similar category interactions, their preferences within those categories may differ, leading to distinct future behaviors. For example, in Fig.~\ref{fig:user_example}, \textit{User 1} and \textit{User 2} share identical category interactions but differ in their preferences, resulting in different subsequent interactions. This is particularly evident with cold users who have a limited interaction history. \\ 
\textbf{Lack of Negative Signals.} We cannot infer category-level negative signals since item-level negative feedback does not imply disinterest in the corresponding categories. \xeg in Fig. ~\ref{fig:user_example}, \textit{User 1} can explicitly mark \textit{Phone 2} as negatives. However, the inferred category \textit{Phones} from this item is not a negative since this user interacts with \textit{Phone 1}. Therefore, unlike at the item level, no explicit negative feedback can be provided at the category level. As a result, negative signals are consistently missing at the category level.\\
\textbf{Precision-oriented Objective.} 
Given the space constraint for CatRec, it's crucial that the top-ranked recommended categories are highly precise. Additionally, due to the limited attention span of customers, it's essential to prioritize only the most relevant categories to capture the user's interest effectively. This makes it unsuitable to adapt item-level recommendations for CatRec directly, as item-level recommendations mainly focus on high recall with a large number of predicted items.



\noindent \textbf{Solution.} To tackle the first challenge, we propose a \textit{user-distinctive category encoder}. As mentioned before, relying solely on category-level information can make users less distinctive, as it overlooks individual preferences within a category. For example, consider two users who both frequently interact with the "Electronics" category. One user might prefer high-end smartphones, while the other leans toward budget laptops. This distinction is evident through their specific item-level interactions within the electronics category. To capture these unique preferences, we leverage these item-level interactions to create a personalized embedding for each category that reflects the unique preferences of each user. Unlike existing methods that produce a single, generic representation for each category across all users, our method creates a distinctive category representation for each user.

To address the second challenge, we design a \textit{probability-weighted negative sampler} to generate strong negative samples by inferring user preferences probabilistically. A desired negative sampler aims to find hard or strong negatives, \ie negatives with high probability to be classified as positives incorrectly, in order to incur high loss for gradient descent.
As there is no explicit negative signals, we must select negatives from those "non-interacted" categories, \ie those have never been interacted positively. We find that the most likely non-interacted categories (i.e., those that are supposed to be most preferred but end up being not interacted with) are also most likely to be strong negatives. Therefore, to rank the non-interacted categories, we train a maximum likelihood estimator (MLE) and generate negatives based on their corresponding probabilities. 
%

To address the last challenge, we apply a \textit{precision-centric loss}, which is proportional to the likelihood of errors for negative samples. Limiting the size of output is equivalent to avoiding false positives, which are categories that are incorrectly predicted to be positive. Therefore, we should prioritize correcting these mistakes by assigning a higher penalty in the loss function based on their likelihood of being predicted. These probabilities have already been generated by our MLE. Therefore, unlike existing works that completely discard the negative sampler after collecting the negatives, we propose to reuse these probabilities to aid preicision optimization.

Experiments show the advantages of proposed models over existing works that treat categories as items and apply item-level recommendations to solve CatRec. The improvement from this model could be at least $8\%$ (relative) compared to the best baseline. In addition, we perform extensive ablation studies to study the influence of each module.
The contribution of this work can be summarized as follows.

\begin{itemize}
\item We construct a  user-distinctive
category encoder to utilize item-level information to generate user-specific category embedding.
\item We propose a novel probability-weighted negative sampler to generate negative samples alongside with their corresponding weight in the fine-grained training. 
\item We design a loss function that is proportional to false positive rates to improve the top outputs from the model.
\item We perform extensive experiments to evaluate the effectiveness of our proposed model and each individual module.
\end{itemize}

The rest of this paper is summarized as follows: In Section 2, we discuss the related works. In Section 3, we discuss the motivation of CatRec. In Section 4, we present an overview of our model, whose details are analyzed in Section 5. In Section 6, experiments are performed to validate the effectiveness of our model and each individual component.

\section{Related Works}
\textbf{Category Recommendation.} Category recommendation is a crucial aspect of personalized recommendation systems, as it helps
users discover a broader range of items that align with their interests and preferences. However,
compared to item-level recommendations \cite{liang2018variational,wu2016collaborative,he2017neural,tay2018latent,wang2019neural,sankar2021protocf,park2012adaptive,yin2012challenging}, category-level recommendation attracts less attention in the
existing literature. The study of \cite{prediction} is the first to investigate how top-level categories can be better predicted in a cold-start setting on eBay by solely exploiting information derived from the social network, i.e., the user’s “likes” from Facebook. However, it is not obvious to assume that every user will have social media interaction, and even if they have, it might be difficult to
obtain that information in most cases. Another work \cite{social} considers using item-level
preferences of users to infer top-level and sub-level purchase category recommendations. However,
achieving category-level recommendations using solely item-level preference makes an assumption that
ignores the differences between item-level recommendations and category-level recommendations. \cite{hierarchical} introduce HUP, a Hierarchical User Profiling framework utilizing Pyramid Recurrent Neural Networks and Behavior-LSTM to model users' real-time interests at multiple scales in E-commerce, incorporating fine-grained sequential interactions for improved hierarchical user profiling and recommendation.

\noindent \textbf{Intention-based Recommendation.} Even though some works do not study category-level recommendation directly, they still predict at the category level to aid item-level prediction. This track calls the item-level preference as intentions. When predicting intentions, existing works \cite{air,intention2,intention3} tend to apply item-level methods directly by replacing items in existing methods with categories. For example, \cite{air} uses an attention-based method to infer category-level interests, similar to the method in an item-level recommender \cite{attention}. Similarly, \cite{intention3} simply feeds categories as items to a Temporal Convolutional Network. Only a few works \cite{intention1} use item-level information to perform predictions on category level. For example, \cite{intention1} proposes to use an attention-based model to generate user-specific embeddings for categories. However, it groups all items from each user together. In this case, even if two interactions, which are correlated to the same category, occur in two different years, it will still group them together and generate the embedding by incorporating the information from both items. Unlike this work, our model uses a VAE to utilize both item-level information and user features to generate embeddings. Our model will group items from the same category if and only if the interactions take place in a short period.

\noindent \textbf{Negative Sampling.}
Negative samples are essential in recommender systems. Existing works \cite{negative5,negative1,negative2,negative3,negative4} primarily focus on identifying strong negatives, while weak negatives, which are infrequently interacted with, contribute minimally to training losses. Some approaches \cite{negative1,negative2} sample from fixed distributions, as seen in \cite{fmlp,sasrec}, often resulting in weak negatives. To generate stronger negatives, methods like \cite{negative3} use scores from previous iterations to select incorrectly predicted items, though these may later become false negatives. Other works \cite{negative4} rank user behaviors and choose weakly interacted items (e.g., viewed but not purchased) as negatives, but even these interactions can be valuable, particularly for new users. \cite{negative5} leverages auxiliary information to generate negatives, though such data isn't always available. \cite{hardness} identifies stronger negatives using hardness, the ratio of scores between positives and negatives in a BPR model \cite{negative1}, but this limits the model to collaborative filtering. In our paper, we propose using a separate $MLE$ model, which doesn't require negatives during training, to guide the selection of negative samples and generate a candidate list. Unlike existing methods that discard intermediate scores after generating negatives, our cascading model introduces an additional loss to reuse these probabilities, thereby reducing false positives.

\section{Motivation}

Online e-commerce nowadays pays attention to CatRec due to its benefits for customers, sellers, and business platforms. The primary goal of CatRec is to guide customers in \textbf{exploration of the e-commerce system}. Unlike item-level recommendations, CatRec focuses on encouraging users to explore various categories rather than prompting immediate transactions.  This deeper exploration increases the likelihood of customers finding products that meet their needs. For existing customers, CatRec engages them by suggesting categories that align with or are similar to their previous interactions. For example, as shown in Fig. ~\ref{fig:user_example}, if User 1 is initially interested in books, they might also be interested in laptops, like User 2. 
Proactively recommending the laptop category to User 1 could diversify their interactions beyond book searches. For new users, category recommendations are vital for navigation, providing a starting point for exploring different areas of interest. This allows them to explore broadly before their preferences are well understood by the recommendation system.

In addition to the benefits of customers, CatRec ensures \textbf{fairness in recommendations}. Traditional item-level recommender systems often favor popular items, leaving lesser-known brands or products overlooked. In contrast, CatRec's category-level predictions reduce this bias, as the popularity gap among categories is comparatively narrower. Moreover, the outputs from CatRec serve as a \textbf{recall set for item-level recommendations}. This structured approach allows businesses to funnel user preferences from broad categories to specific items, making the final recommendations more tailored and likely to result in a purchase. In addition, feedback at the category level often indicates "\textbf{user preferences}" more than feedback on specific items. Users may not like a particular item but still show interest in the general category, providing valuable insights for future recommendations.


Finally, CatRec plays a crucial role in \textbf{business decisions and platform campaigns} by enabling effective and organized recommendations. E-commerce platforms use various channels for recommendations, such as text messages, pop-up ads, and emails. Techniques like email recommendations have limited space, making it more practical to suggest broader categories instead of specific items. Eg., highlighting three categories rather than three individual products creates a cleaner and more appealing presentation that captures user interest without overwhelming them. Additionally, CatRec helps identify and filter target users for specific marketing campaigns. By understanding which categories attract certain users, platforms can promote products more effectively. Eg., if an online platform wants to promote a new product in the cell phone category, CatRec can help identify the target audience and implement promotion strategies like coupons and discounts, even without prior interactions with the product.

In an anonymous e-commerce platform, CatRec has already \textit{demonstrated its merits in practice}. It allows the deployment of a message ranking model to prioritize messages from various sources based on user relevance. This addresses the issue of users receiving unorganized message bombardment, leading to significant success in production. The message consists of a bunch of similar items, each message can be regarded as a granular category. It achieved 8\% increase in quality visits, 6\% increase in click-through rate, and 18\% reduction in opt-out rate in the market. Furthermore, a stratified model to understand and enhance users’ long-term interests was built upon category recommendation. It incorporates insights from various perspectives, including categories users interacted with, similar categories, and non-linearly connected categories, aiming to capture user interests comprehensively. It results in a 30\% increase in Click-Through Rate.

\section{Model Overview}
In this section, we first formally introduce the notations and the problem in this paper. Then, we provide a high-level overview of our proposed model.

\subsection{Problem Formulation}
\begin{table}[]
\begin{tabular}{ll}
\hline
Notation & Descriptions     \\ \hline
$u_t$       & User $t$           \\
$i_b$       & Item $b$           \\
$c_x$ & Category $x$ \\
$g$ & A function maps an item to a category\\
$f_t$ & The feature of user $t$\\
$\pi_t$ & The sequence of past interacted items of user $t$\\
$\delta_t$ & The sequence of past  interacted categories of user $t$\\
$\gamma_t$ & The sequence of future interacted categories of user $t$\\
$k$ & The maximum length of $\delta_t$\\
$M_1$ & The MLE model \\
$r_t$ & The category list generated by $MLE$ for user $t$\\
$M_2$ & The final prediction model\\
\hline
\end{tabular}
\caption{Notations used in this paper.}
\label{notation}
\vspace{-8mm}
\end{table}

Let $U$ = $\{u_1,u_2,...,u_n\}$ be the set of users and $I$ = $\{i_1,i_2,...,i_m\}$ be the set of items. Given a set of categories $C=\{c_1,c_2,...,c_s\}$, we have a mapping function $g:I\mapsto C$ indicating the category of each item. \xeg $g(\text{Book 1})=$ Book. Each user has some known features $F$=$\{f_1,f_2,...,f_n\}$.
Let $\Pi$=$\{\pi_1,\pi_2,...,\pi_n\}$ be the set of the sequence of past interactions between users and items. In other words, for each user $u_j$, we have a sequence of past interacted items $\pi_j \subseteq I$. \xeg $\pi_1$=\{Phone 1, Office 1\} in Fig. ~\ref{fig:user_example}.
The users' corresponding category-level sequence is then denoted by $\delta_t=\{g(i_x)|i_x \in \pi_t\}\subseteq C$. \xeg $\delta_1=\{\text{Phone},\text{Office}\}$. Let $k$ be the maximum length of category sequences. For an arbitrary user $u_t$, given this user’s feature $f_t$ and past interactions at item level $\pi_t$ and category level $\delta_t$, we aim to recommend top $N$ categories $\Gamma$ =$\{\gamma_1,\gamma_2,...,\gamma_N\}$ that the users are most likely to interact with in the future, \eg $\gamma_1$=\{book\} for $u_1$. The summary of notations is shown in Table.\ref{notation}.

\subsection{Model Overview}
\begin{figure}[h]
\centering
\includegraphics[width=0.9\linewidth]{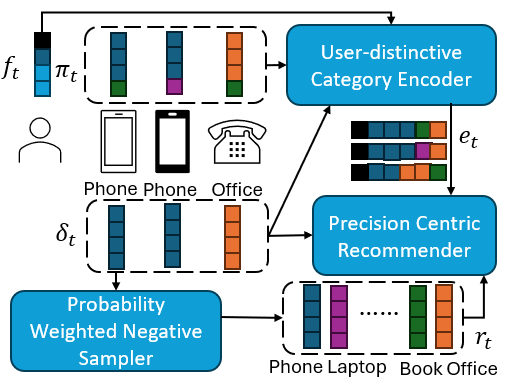}
\caption{Model Overview}
\label{fig:overview}
\end{figure}

As discussed in Section \ref{sec:intro}, we have three challenges and build three modules to address them. We show the three components of our proposed framework in Fig. ~\ref{fig:overview}.

The first part, the Probability Weighted Negative Sampler, appears on the lower left. This module aims to provide us with strong negative samples that are likely to be predicted as positives. Therefore, for a a strong negative category X, Pr(X is predicted and X is negative) must be high. This expression can be rewritten as $$\Pr(\text{X is predicted})\Pr(\text{X is negative}|\text{X is predicted})$$ 
We can infer the first part, $\Pr(\text{X is predicted})$, using an MLE. Then the non-positives in the training datasets can be utilized to identify the real negatives. In this case, categories that are not positives can serve as negatives. Then our MLE provides us with a ranking of these negatives based on the above expression. We follow \cite{classification}, which views the task of candidate generation as a classification problem. As we discussed in Section \ref{sec:intro}, the item-level negative is inappropriate for category-level negative samplings, so we ignore the item-level interactions ($\pi_t$) when building the first module.

The choice of negative samples may appear too assertive, yet within the context of category recommendation, their use carries a low risk of being false negatives. Selecting highly likely items from an MLE model can lead to high false negative rates due to the potential similarity between items and the fluid nature of user interests. In contrast, categories are inherently dissimilar, as otherwise, they would have already been merged into a broader category. Given the apparent distinctions between categories, transitions among them are much harder. Leveraging these substantial gaps within categories enables us to select stronger negative samples, contributing to improved model fine-tuning.

Next, we build the User-distinctive category encoder to train the embedding for each category interaction of each user, which is later used in our prediction model, as shown in the uppermost part in Fig. ~\ref{fig:overview}. The primary objective of this approach is to discern between users who exhibit similarities at the category level. To achieve this, we leverage a combination of a user's item-level interactions within each category and their demographic information as input for the VAE. The VAE is designed to reconstruct this input information, with the output from the encoder serving as the embedding for each category interacted with by a specific user. Consequently, this yields distinct embeddings for each category, dependent on both the user and the items within the category.

The final prediction model shown in the lower right corner of Fig. ~\ref{fig:overview} is a simple MLP akin to conventional approaches. However, we have customized the loss function to leverage insights from our negative sampler. As discussed in Section \ref{sec:intro}, our model can only generate a short list of recommended categories. Therefore, we emphasize minimizing false positives. To achieve this, our cascaded model primarily focuses on optimizing precision rather than recall, distinguishing it from item-level recommenders. We have designed a specialized differentiable loss function that adjusts penalties based on the preceding module's output. We want this fine prediction model to avoid the mistakes made by MLE as much as possible, which are incorrect categories that receive high probability scores. 

\section{Details of Model}
In this Section, we present the details of the modules in our work.
\subsection{Probability Weighted Negative Sampler}
\begin{figure}[h]
\centering
\includegraphics[width=\linewidth]{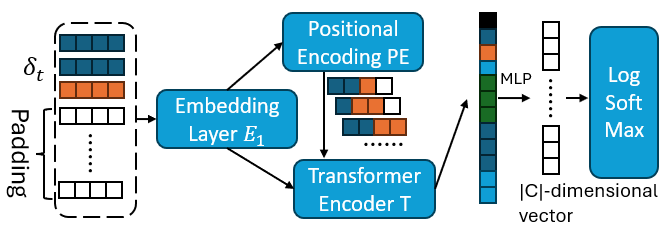}
\caption{Detail of Probability Weighted Negative Sampler}
\label{fig:mle}
\end{figure}

Our MLE-based sampler $M_1$ consists of a transformer-based sequence encoder, an embedding layer, two fully connected layers, and a LogSoftMax layer, as shown in Fig. ~\ref{fig:mle}.

As our first step, our model takes the sequence of past topics $\delta_t$ as input to an embedding layer to get a representation of categories. Suppose $d_1$ is the embedding dimension of our MLE, then our embedding sequence is $E_1(\delta_t) \in \mathbb{R}^{k \times d_1}$ as k is the maximum length of interaction sequence. Next, with the encoding of each category, we generate the embedding for the entire sequence using transformer like existing works \cite{bst}. In addition to the original encoding of categories, the positional information is also learned through multi-head attention. We use the same positional encoding $(PE)$ function from \cite{transformer}. With the output from positional encoding $PE(E_1(\delta_t))$, we get the sequence encoding from the transformer $T$. Then, we train a simple MLP with the output from the transformer, $T(PE(E_1(\delta_t)),E_1(\delta_t))$. A two-level Fully Connected (FC) network gives us a probability vector of dimension $|C|$. Finally, a LogSoftMax layer will utilize this probability vector to give us the final output.

Since the number of categories is often much smaller compared to items, we can treat this recommendation problem as a classification problem like \cite{classification}. Each user will be classified into one of the $|C|$ classes, which indicates the user's future preference in the corresponding categories. Therefore, we use the negative log-likelihood loss (NLLLoss), a loss function frequently used for classification. Then, our loss function for the MLE model, $M_1$, is
$$L_{MLE}=\sum_{\delta_t \in \Delta} \sum_{c_j \in \gamma_t} NLLLoss(M_1(\delta_t),c_j)$$
The categories with top probabilities will form a list of $r_t$, which will be used in our final prediction model. During the training, in addition to this list of categories, we also pass their corresponding probability vector $y_t^{M_1}$ to our final prediction model. This is used to compute the loss for false positives, as we will see in Section 5.3.

\subsection{User-distinctive Category Encoder}

As mentioned earlier, one major challenge is that not all users like the same kinds of products within a category. This means that even if two users, called $\mathbf{A}$ and $\mathbf{B}$, both interact with a product category called Category T, they might pick totally different items from that category. This difference in choices highlights a big problem with just using category-level interaction to understand what users like; it misses the specific details of each user's preferences. To solve this problem, our method focuses on looking closely at the items users interact with within certain categories. By doing so, we can create a better representation (i.e., embedding) of a category that takes into account these finer details of what different users prefer.

However, this brings us to the question of how to effectively integrate this detailed item-level information with the broader category-level data to create a meaningful category representation. A straightforward solution might involve directly concatenating the category identifier with an item-interaction vector to form a category embedding. While this approach is simple and intuitive, it has notable drawbacks. First, this direct aggregation might fail to capture the complex and often non-linear relationships between a category and the specific items within it. The interaction between a category and the items it contains is not merely additive; it involves intricate patterns that a simple concatenation may not effectively represent. Second, concatenating category and item-level information could lead to an embedding that is excessively high-dimensional. Such a large vector could be computationally inefficient, making it cumbersome to process and potentially leading to overfitting.

\begin{figure}[h]
\centering
\includegraphics[width=0.8\columnwidth]{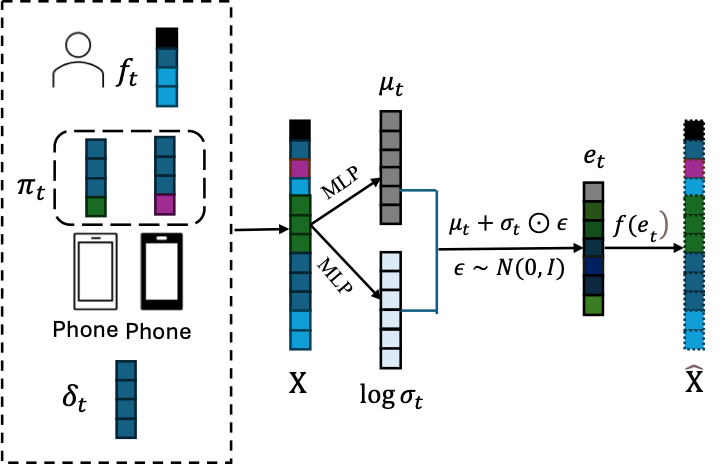}
\caption{Detail of User-distinctive Category Encoder}
\label{fig:topic_emb}
\end{figure}

To address these drawbacks, we use a Variational Autoencoder (VAE) \cite{kingma2013auto} to create a concise embedding (as shown in Fig ~\ref{fig:topic_emb}). VAEs excel at learning complex data distributions, making them ideal for capturing the relationships between a user's interactions with different categories \cite{liang2018variational}. The probabilistic nature of VAEs also enables the generation of smooth, continuous embeddings that can reflect the subtle preferences of users. Additionally, the reconstruction objective of the VAE helps maintain a strong connection to the original data, preserving user preferences while uncovering hidden patterns.

For a given user $t$, our model inputs a concatenated vector $\mathbf{X}=[f_t; \pi_t; \delta_t]$, consisting of $f_t$, the demographic features of the user; $\pi_t$, the user's item-interaction profile within a particular category; and $\delta_t$, the representation of the category with which the user has interacted. This vector is processed through two Multilayer Perceptron (MLP) units, producing two outputs: the mean $\mu_t$ and the log-variance $\log \sigma_t^2$ of a latent distribution. By sampling from this distribution using the parameters $\mu_t$ and $\log \sigma_t^2$, we obtain a hidden representation $e_t$, which is then decoded to reconstruct the input into $\mathbf{\hat{X}}$. This process ensures that the resultant embedding, $e_t$, accurately encapsulates the item-level preferences that are tailored to user demographics within the context of the interacted category.

\subsection{Precision Centric Recommender}
Our prediction model is a simple MLP model with a loss function utilizing the output from our MLE. The loss function consists of two parts -- a penalty based on false positives from MLE and an error for false negatives from itself.

\begin{figure}[h]
\centering
\includegraphics[width=\linewidth]{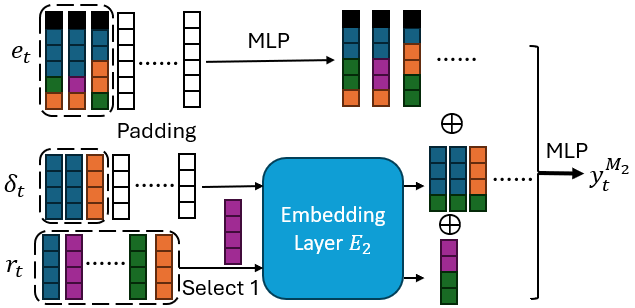}
\caption{Detail of Precision Centric Recommender}
\label{fig:prediction}
\end{figure}

Our last module, as shown in Fig.~\ref{fig:prediction}, uses the outputs from the previous two modules to perform a fine prediction. It takes the embedding from VAE $e_t$ as pretrained embedding and uses an MLP to process them. Also, one more embedding layer $E_2$ will learn the embedding for past categories and each individual category $c_j \in r_t$. All these embeddings will be concatenated together as input to an MLP to give us the score for category $c_j$ for this user.

As we discussed in Section 4, we expect the model to correct the mistakes made by our MLE and avoid false positives. Therefore, we design a loss function that results in a higher loss for categories with a high likelihood in $MLE$. On the other hand, if the output from MLE is indeed a true positive, it should incur zero loss. Such piecewise functions should be differentiable for training purposes. Therefore, we design the following loss function for the MLE. 
$$L_{precision} = \sum_t (ReLU(y_t^{M_1}-y_{t}))^2$$
Here, $y_t \in \{0,1\}^{|r_t|}$ is a sequence indicating the ground truth of each category in $r_t$. Its $j^{th}$ element is 1 if and only if $c_j \in \gamma_t$.

In addition, the model also needs to penalize false negatives. We use the standard mean-square error as our loss, which gives us
$$L_{MSE}=\sum_{t}(y_t-y_t^{M_2})^2$$

As a result, the loss for our precision-centric recommender is
$$L_{total} = L_{precision} + L_{MSE}$$

\section{Experiment}
In this section, we perform extensive experiments to evaluate the effectiveness of our proposed model and the contribution of each individual module in CatRec on its own. We provide our implementation and code in \url{https://anonymous.4open.science/r/CCRec-AF16/}
\subsection{Dataset}

We evaluate our model using three datasets: one from an anonymous industry source and two publicly available datasets, RetailRocket \cite{rrdata} and Tmall \cite{Tmall}. Table \ref{dataset} provides details on these datasets. The industry dataset includes interactions from 103,189 users on an e-commerce website, spanning February 1 to April 15, 2023. We designate March 1, 2023, as the cutoff date, categorizing interactions before this date as past interactions and those after as future interactions. On average, each user has 6.43 known interactions and 3.94 interacted categories, with mixed-activity users for training and fewer interactions (1.53 known categories on average) for testing. We also have demographic data, including age group, parental status, and gender. For RetailRocket, we group interactions from the same category within 5 days as one interaction and limit users to a maximum of 2 known interactions to simulate a cold-start scenario. For the Tmall dataset, we use a similar 5-day grouping but limit each user to 10 known interactions and 3 future interactions due to the larger volume of data.

\begin{table*}[h]
\begin{tabular}{c|cccccc}
\hline
             & \# of users & \# of items & \# of categories & \# of past actions & \# of past categories/user & \# of future categories/user \\ \hline
Industry (Train) & 103189           &       3207     & 151              & $6.43 \pm 7.72$             & $3.94 \pm 3.66 $                      & $3.93 \pm 3.70$                         \\
RetailRocket (Train) & 211105           & 91145           & 1107             & $2.21 \pm 1.87$            & $1.39 \pm 0.49 $                      & $ 2.23 \pm 6.27$                         \\ 
Tmall (Train) & 375407           & 2353207           & 72                         & $11.52 \pm 13.63 $  & $2.77 \pm 4.71$                    & $ 2.72 \pm 0.50$                         \\ \hline
Industry (Test) & 222           & 3207           & 151             & $2.10 \pm 5.08$             & $1.53 \pm 1.23$                      & $1.45 \pm 1.17$ \\
RetailRocket (Test) & 23457           & 91145           & 1107 &    $2.22 \pm 1.85 $          & $ 1.33 \pm 0.47 $                     & $ 1.51 \pm 3.84 $    \\
Tmall (Test) & 41712           & 2353207           & 72      & $7.10 \pm 10.73 $       & $2.87 \pm 2.13$                                  & $ 2.45 \pm 0.69$         
\\ \hline

\end{tabular}
\caption{Summary of Dataset}
\label{dataset}
\end{table*}

\noindent \textbf{Implementation Detail.}
We implemented our model using pytorch 1.9.0 on a machine running Ubuntu 18.04. All the MLPs in our model have 2 layers, and the embedding layers have a dimension of 64. The output dimension of MLE for each dataset is equal to its number of categories. The output dimension of our VAE is 256. We select the values of $k$ based on the average number of categories as shown in Table.~\ref{dataset}. For the industry and Tmall dataset, $k=10$. Since the average number of interactions was much smaller in the RetailRocket dataset, we set $k=5$. The learning rates for all the models were 0.0001, and we trained all the models for 100 epochs.

\subsection{Baselines}
Our research problem focuses on category recommendation in a cold start setting. This scenario cannot be perfectly matched in existing works. Therefore, we select three different types of baselines -- recommendation baseline, cold-start baseline, and VAE-based baseline. For recommendation baselines and cold-start baselines, we directly use categories as items in the original model.

\begin{itemize}
\item \emph{FMLPRec} \cite{fmlp}. An MLP-based model with an enhanced filter to handle potential noise in the data.
\item \emph{SASRec} \cite{sasrec}. A self-attentive model based on MLP to achieve decent performance in both dense and sparse datasets.
\item \emph{CLRec} \cite{contrastive}. A contrastive learning approach, originally designed for item-level recommendations.
\item \emph{Mamba4Rec} \cite{mamba}. A recommender system utilizing Mamba to encode sequential information. 
\item \emph{MeLU} \cite{melu}. A meta-learning method that trains a global model and then optimizes it separately for each cold user. 
\item \emph{VAERec} \cite{wu2020hybrid}. This VAE-based \cite{kingma2013auto} model was previously used as an item recommendation \cite{wu2020hybrid}. 
\item \emph{VAERec+}. A modification from VAERec adds the user's demographic information.
\item \emph{TIGER} \cite{tiger}. A generative retrieval-based solution that leverages RQ-VAE to generate semantic IDs for recommendation.
\end{itemize}
We used a learning rate of 0.0001 for our model and all the baselines. For each baseline, we reused the same architecture and other hyperparameters as the source codes provided by the authors.

\subsection{Result Analysis}
To evaluate the performance of our proposed models, we adopt several standard evaluation metrics, including Hit Ratio (HR), precision, recall, and F1-score. In the industry dataset, we evaluate our model on two different sets- warm users and cold users. We sample $10\%$ of warm users from the training set as the testing set. In addition, we use 222 cold users, as shown in Table.~\ref{dataset}. Even though the prediction on categories is often needed when the users are cold, we can still utilize category-level prediction for warm users \cite{air}. The performance for warm users and cold users in the industry dataset is shown in Table.~\ref{industrywarm} and \ref{industrycold}, respectively. We also tested on the RetailRocket and Tmall datasets, and the results are shown in Appendix A. The best results among baselines are underlined, and the best overall results are highlighted.

\iflong
\begin{table*}
\begin{tabular}{c|ccccccccccc}
\hline
 & HR@1 & HR@3 & HR@5 & Precision@3 & Precision@5 & Recall@1 & Recall@3 & Recall@5 & F1@1 & F1@3 & F1@5 \\ \hline
FMLPRec & 0.0132 &0.0403 &0.0672 &0.0135 &0.0137 &0.0084 &0.0257 &0.0434 &0.0102 &0.0177 &0.0208          \\
SASRec  & 0.0119 &0.0388 &0.0685 &0.0131 &0.0139 &0.0075 &0.0249 &0.0442 &0.0092 &0.0172 &0.0212       \\
VAERec &   0.2806 & 0.4245 & 0.5087 & 0.1540 & 0.1165 & 0.2048 & 0.3212 & 0.3953 & 0.2368 & 0.2082 & 0.1799           \\
VAERec+  & 0.2377 & 0.3975 & 0.4835 & 0.1445 & 0.1103 & 0.1697 & 0.2966 & 0.3726 & 0.1981 & 0.1943 & 0.1702           \\
MeLU &  0.1289 &0.2638 &0.3503 &0.0929 &0.0767 &0.0817 &0.1766 &0.2431 &0.1 &0.1217 &0.1166\\
CLRec & 0.1645 & 0.3165 & 0.3924 & 0.1055 & 0.0784 & 0.1582 & 0.3101 & 0.3861 & 0.1613 &  0.1574 & 0.1304 \\
TIGER  & \underline{0.3205}   & \underline{0.4681}  & \underline{0.5345}   & \underline{0.1707}   & \underline{0.1214}   & 0.2379  & 0.3565  & 0.4159  & 0.2731  & \underline{0.2308}  & \underline{0.1880}  \\
Mamba4Rec &   0.2238 & 0.3507 & 0.4141 & 0.1237 & 0.0905 & \textbf{\underline{0.3914}} & \textbf{\underline{0.5161}} & \textbf{\underline{0.5654}} & \textbf{\underline{0.2847}} & 0.1995 &0.1560    \\
 \hline
CCRec   &   \textbf{0.3360} & \textbf{0.4885} & \textbf{0.5560} & \textbf{0.1803} & \textbf{0.1285} & 0.2129 & 0.3427 & 0.4073 & 0.2606 & \textbf{0.2363} & \textbf{0.1954}            \\ \hline
\end{tabular}
\caption{Performance of Warm Users in Industry Dataset}
\label{industrywarm}
\end{table*}

\begin{table*}
\begin{tabular}{c|ccccccccccc}
\hline
        & HR@1 & HR@3 & HR@5 & Precision@3 & Precision@5 & Recall@1 & Recall@3 & Recall@5 & F1@1 & F1@3 & F1@5 \\ \hline
FMLPRec &   0.0045 &0.0315 &0.0675 &0.012 &0.0144 &0.003 &0.0247 &0.0495 &0.0036 &0.0161 &0.0223        \\
SASRec  &  0.0225 &0.0495 &0.0765 &0.0165 &0.0153 &0.0154 &0.034 &0.0526 &0.0183 &0.0222 &0.0237          \\
VAERec &   0.3739 & 0.5045 &  0.6036 & 0.1772 &  0.1315 &  0.3081 & {0.4190} & {0.5090} & 0.3378 & \underline{0.2490} & \underline{0.2090}            \\
VAERec+  & 0.2883 & 0.4460 & 0.5225 & 0.1577 & 0.1198 & 0.2259 & 0.3560 & 0.4403 & 0.2533 &  0.2185 & 0.1884           \\
MeLU & 0.1081 &0.2432 &0.3063 &0.087 &0.0666 &0.0743 &0.1795 &0.2291 &0.088 &0.1172 &0.1032 \\
CLRec & 0.1595 & 0.3298 & 0.3936 & 0.1099 & 0.0787 & 0.1596 & 0.3298 & 0.3851 & 0.1596 &  0.1649 & 0.1307 \\
TIGER & \underline{0.4729}  & \underline{0.5800}  & \underline{0.6081} & \underline{0.2132}  & \underline{0.1369}  & 0.3934 & 0.4933 & 0.5156  &  \underline{0.3695} & 0.2907   & 0.2164  \\
Mamba4Rec &   0.2805 & 0.4118 & 0.4796 & 0.1463 & 0.1032 & \textbf{\underline{0.4882}} & \textbf{\underline{0.5988}} & \textbf{\underline{0.6368}} & 0.3563 & 0.2352 & 0.1776 \\
 \hline
CCRec   &   \textbf{0.4819} &\textbf{0.5855} & \textbf{0.6261} &\textbf{0.2162} & \textbf{0.1414} & {0.3312} & {0.4458} & {0.486} & \textbf{0.3926} & \textbf{0.2912} & \textbf{0.2191}           \\ \hline
\end{tabular}
\caption{Performance of Cold Users in Industry Dataset}
\label{industrycold}
\end{table*}

\fi

\iflong
\begin{table*}
\begin{tabular}{c|ccccccccccc}
\hline
        & HR@1 & HR@3 & HR@5 & Precision@3 & Precision@5 & Recall@1 & Recall@3 & Recall@5 & F1@1 & F1@3 & F1@5 \\ \hline
MLE &   0.1929 &0.3401 &0.4196 &0.1226 &0.0943 &0.1222 &0.2331 &0.2989 &0.1496 &0.1607 &0.1434            \\
MLE+VAE  & 0.3197 &0.4624 &0.5373 &0.1686 &0.1231 &0.2025 &0.3205 &0.3899 &0.248 &0.221 &0.1871            \\
MLE+Cascading  & 0.3377 &0.489 &0.5557 &0.1802 &0.1284 &0.2139 &0.3425 &0.4067 &0.2619 &0.2362 &0.1951           \\ \hline
CCRec   &    0.336 &0.4885 &0.556 &0.1803 &0.1285 &0.2129 &0.3427 &0.4073 &0.2606 &0.2363 &0.1954             \\ \hline
\end{tabular}
\caption{Ablation Study in Industry Dataset for Warm Users}
\label{industryabl_warm}
\end{table*}

\begin{table*}
\begin{tabular}{c|ccccccccccc}
\hline
        & HR@1 & HR@3 & HR@5 & Precision@3 & Precision@5 & Recall@1 & Recall@3 & Recall@5 & F1@1 & F1@3 & F1@5 \\ \hline
MLE & 0.2342 &0.3558 &0.4189 &0.1291 &0.0945 &0.1609 &0.2662 &0.325 &0.1908 &0.1739 &0.1465    \\
MLE+VAE  & 0.4324 &0.572 &0.6306 &0.2057 &0.1387 &0.2972 &0.4241 &0.4767 &0.3522 &0.277 &0.2149 \\
MLE+Cascading    &   0.4594 &0.5855 &0.6261 &0.2147 &0.1423 &0.3157 &0.4427 &0.4891 &0.3743 &0.2891 &0.2205          \\ \hline
CCRec   &  0.4819 &0.5855 &0.6261 &0.2162 &0.1414 &0.3312 &0.4458 &0.486 &0.3926 &0.2912 &0.2191         \\ \hline
\end{tabular}
\caption{Ablation Study in Industry Dataset for Cold Users}
\label{industryabl_cold}
\end{table*}
\fi

For warm users from the industry dataset, we can see TIGER and Mamba4Rec are the best baselines in most of the cases. The differences between models are small when the output sizes are large. This is expected due to the small number of categories. In our industry dataset, we only have 151 categories. Therefore, if the model can output more than 5 categories, the task becomes much easier such that every model can achieve decent performance.

Our proposed model can outperform all baselines in terms of precision and hit ratio. If we only replace items with categories using existing item-level recommenders, information related to the underlying differences at item levels will be ignored. They also overlooked the differences in problem space and chose conservative negative samples. \xeg FMLPRec, and SASRec use 99 randomly sampled negatives. Due to the difficulty in transferring interests at the category level, most of the negatives have already received low scores in prediction. However, the improvement of our model is also small when the output size is large. In real-world applications, we do not need a lot of recommended categories. The users can engage with a sufficient amount of items by knowing a few categories. 

Since our model mainly focuses on optimizing precision, our recall scores are relatively lower. Some baselines, like Mamba4Rec, mainly aim to improve the recall of recommended categories. Therefore, their recall scores become higher compared to our framework. However, when taking both recall and precision into consideration to compute the F1 scores, our model can still outperform all baselines in almost all cases.

When testing cold users in our industry dataset, our model demonstrates similar improvements. Note that the results in Table.\ref{industryabl_cold} and Table.\ref{industryabl_warm} are not directly comparable since they are tested using different testing sets. The VAE module in our framework further improves the quality of recommended categories. As we will see in Section 6.4, VAE is more influential for cold settings.

\subsection{Model Analysis}
In this section, we study the role of each individual module through an ablation study. 
Also, the influence of hyper-parameters used in our model will be analyzed through extensive experiments.

\subsubsection{Ablation Study}
In our model, we have three different modules, as we discussed in Section 4. To evaluate them separately, we build four different models.
\begin{itemize}
\item  MLE. We use the MLE=Relevant Category generator to perform prediction directly.
\item  MLE+VAE. Instead of feeding the category-level information directly to MLE, we use VAE output to provide a pretrained category embedding for each user.
\item  MLE+Cascading. We simply use our cascading model without introducing additional embedding.
\end{itemize}

The performance of these models for warm users and cold users are presented in Table.\ref{industryabl_warm} and \ref{industryabl_cold}, respectively. As expected, the MLE has the worst performance, so we picked it as the baseline to explore the potential improvement from each module. We first look at the warm case. Overall, our VAE can give us around $10\%$ improvement for the three hit ratio metrics. The improvement from our cascading structure is slightly higher than the MLE, which is also around $10\%$. In terms of warm users, we find that our cascading structure will be more useful than the VAE model.

If we compare our cascading model against our final model, we find that its performance will be very close to our best model. In other words, for our warm users, we find that VAE will have little influence. Such observation is expected because our warm users can be easily distinguished by their distinct past history. As a result, we do not need item-level information to enrich the embedding and tell the differences among these warm users.

The improvement of our model over MLE becomes much smaller when our output size is larger, which shows similar patterns to our comparisons against baselines. This is also expected because the task again becomes much easier when we can output 5 or 10 categories. Since the task is no longer very challenging, even a simple model like MLE can give us decent results.

The improvement of modules becomes slightly different for cold users. For warm users, VAE does not give us much improvement because we can easily distinguish those users. However, when the testing users, on average, only have 1.5 known interacted categories, we need the item-level information to tell the differences among users and give us more information to help prediction. Cascading architecture is still slightly more important than our VAE. However, when combining them together, unlike the warm case, our model can achieve better results than a model without VAE.

\iftrue
\subsubsection{Parameter Analysis}
One of the key parameters in our model is related to handling the negative samples. The amount of negative samples, which is controlled by $|r_t|$, has a significant impact on our performance. Therefore, we vary the value of $|r_t|$ and use $HR@1$ and $HR@3$ as evaluation metrics. The results are plotted in Fig. ~\ref{fig:Parameter Analysis}.

\iflong
\begin{figure}[h]
\centering
\includegraphics[width=\linewidth]{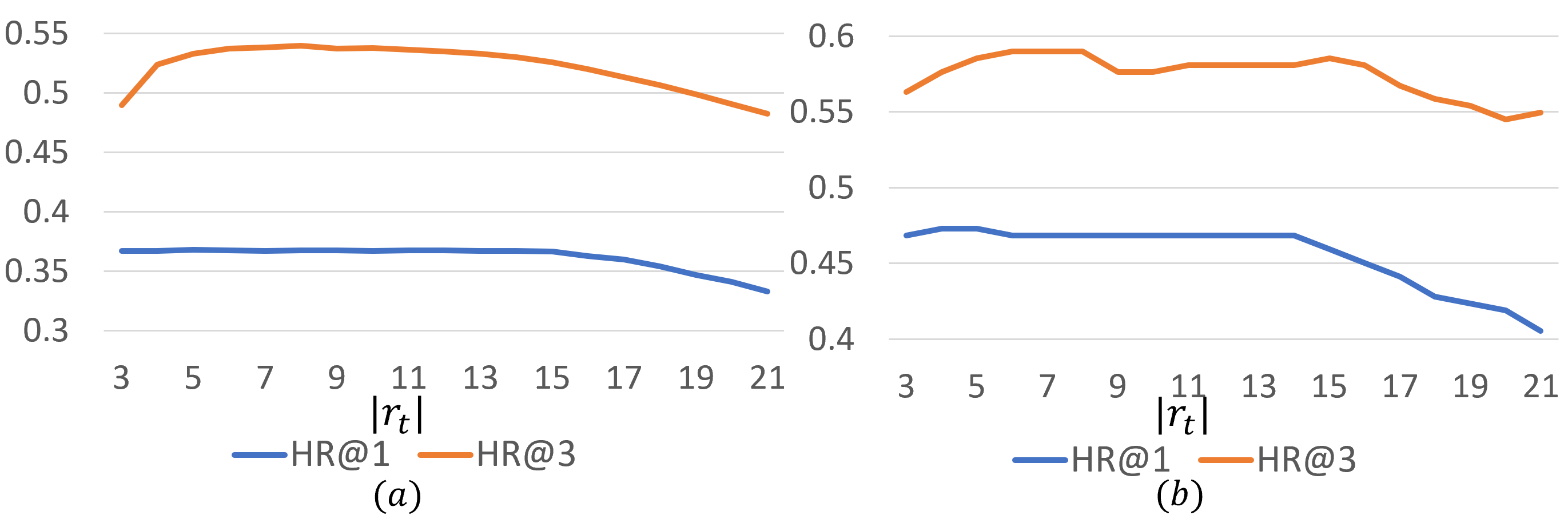}
\caption{Effect of $|r_t|$ in Industry Dataset for (a) Warm Users and (b) Cold Users}
\label{fig:Parameter Analysis}
\end{figure}
\fi

For both cold and warm cases, we can see the value of $|r_t|$ has little influence on $HR@1$ when $|r_t| \leq 11$. As we further increase $|r_t|$, $HR@1$ begins to drop. Both observations are expected. $HR@1$ should not change much because only the highest-ranked category will influence the result. Our model will penalize more for categories that receive high scores when it is a false positive from the first model. Therefore, for categories that still get the highest score from our model, it means that it is unlikely to be false positive. Therefore, our model achieves high confidence for the output of the highest score. Then, when the output size gets larger, the additional negative samples will have little influence on weights. On the other hand, it will hurt the model by introducing more noise. Therefore, the performance will slightly drop when $|r_t|$ gets too large.

The pattern of $HR@3$ is slightly different. We can observe an improvement of $HR@3$ at an early stage. This is caused by initially, the ground truth may not be covered in $r_t$. As a result, the model cannot find the correct categories if they are not in the input. However, when $|r_t|$ gets larger, we have the same observations due to the aforementioned reasons. The improvement is more obvious for cold users than warm users because, for cold users, there are fewer ground truths. As a result, small $r_t$ is more likely to miss the ground truth. As a result, increasing its size is more effective for cold users.

In conclusion, if we want the output size to be small, we should also pick a relatively small $|r_t|$. If we want to get a larger output size, then we may need to slightly increase $|r_t|$ to get better performance.
\fi
\section{Conclusion}
In this work, we aim to address a relatively new problem: category-level recommendation. We propose a cascading model to provide negative samples with adaptive loss. To distinguish users at category levels and introduce more information to aid prediction, we add a VAE to help us generate item-level dependent category embedding. By taking advantage of these two modules, our model outperforms existing works by a large margin.

\bibliographystyle{ACM-Reference-Format}
\bibliography{reference} 

\begin{table*}[ht]
\begin{tabular}{c|ccccccccccc}
\hline
        & HR@1 & HR@3 & HR@5 & Precision@3 & Precision@5 & Recall@1 & Recall@3 & Recall@5 & F1@1 & F1@3 & F1@5 \\ \hline
FMLPRec    &   0.0013 &0.0049 &0.0076 &0.0016 &0.0015 &0.0004 &0.0018 &0.0029 &0.0007 &0.0017 &0.002      \\
SASRec &  0.0015 &0.0033 &0.0054 &0.0011 &0.0011 &0.0005 &0.0012 &0.002 &0.0008 &0.0011 &0.0014  \\
 VAERec &  0.1278 & 0.2757 & 0.3668 & 0.1062 & 0.0918 & 0.0798 & 0.1889 & 0.2648 & 0.0982 & 0.1360 & 0.1364 \\
CLRec & 0.0348 & 0.0932 & 0.1302  & 0.0317 & 0.0266 & 0.0321  & 0.0864 & 0.1199  & 0.0334  & 0.0464  & 0.043 \\
TIGER & 0.1997 & 0.3477 & 0.4247 & 0.1327 & 0.1051 & 0.1360 & 0.2497 & 0.3175& 0.1618 & 0.1733 & 0.1579 \\
Mamba4Rec &   \underline{0.6068} & \underline{0.7081} & \underline{0.7298} & \underline{0.2431} & \underline{0.1513} & \textbf{\underline{0.6129}} & \textbf{\underline{0.6555}} & \textbf{\underline{0.6638}} & \textbf{\underline{0.6098}} & \underline{0.3546} & \underline{0.2464} \\
           \hline
CCRec   &     \textbf{0.6844} & \textbf{0.7641} & \textbf{0.7722} & \textbf{0.2634} & \textbf{0.1604} & 0.5208 & 0.6177 & 0.6282 & 0.5915 & \textbf{0.3694} & \textbf{0.2555}             \\ \hline
\end{tabular}
\caption{Performance in RetailRocket Dataset}
\label{rrcold}
\end{table*}

\begin{table*}[ht]
\begin{tabular}{c|ccccccccccc}
\hline
        & HR@1 & HR@3 & HR@5 & Precision@3 & Precision@5 & Recall@1 & Recall@3 & Recall@5 & F1@1 & F1@3 & F1@5 \\ \hline
FMLPRec    &   0.0094 &0.0169 &0.0426 &0.0057 &0.0086 &0.0041 &0.0073 &0.0183 &0.0057 &0.0064 &0.0117      \\
SASRec &  0.0097 &0.0183 &0.0433 & 0.0061 &0.0087 &0.0042 &0.0079 &0.0186 &0.0058 &0.0069 &0.0119  \\
 VAERec &  0.0001 & 0.1764 & 0.5040 & 0.0588 & 0.1088 & 0.0001 & 0.0994 & 0.2712 & 0.0002 & 0.0739 & 0.1553 \\
CLRec & 0.0788 & 0.0788 & 0.1343 & 0.0263 & 0.0282 & 0.0320 & 0.0321 &0.0555 & 0.0456 & 0.0289 & 0.0373 \\
TIGER & \underline{\textbf{0.3244}}  & \underline{0.3938} & \underline{0.5664}  & \underline{0.1349}  & \underline{0.1247}  & 0.1081  & 0.1349 & 0.2079 & \underline{0.1622}  & 0.1348  & 0.1560 \\
 Mamba4Rec &  0.0396 & 0.3035 & 0.5263 & 0.1021 & 0.1135 & \underline{\textbf{0.2069}} & \textbf{\underline{0.6685}} & \textbf{\underline{0.7888}} & 0.0665 & \underline{0.1771} & \underline{0.1984} \\
           \hline
CCRec   &     0.2679 & \textbf{0.5239} & \textbf{0.6846} & \textbf{0.1995} & \textbf{0.1799} & 0.1321 & 0.2536 & 0.3382 & \textbf{0.1769} & \textbf{0.2233} & \textbf{0.2349}             \\ \hline
\end{tabular}
\caption{Performance in Tmall Dataset}
\label{tmallresult}
\end{table*}

\appendix
\newpage
\section{Additional Results}
Here we provide the experiment results in RetailRocket dataset and Tmall dataset in Table.~\ref{rrcold} and \ref{tmallresult}, respectively.

We have similar observations as we did in the industry dataset. Since there is no demographic information in these datasets, VAERec+ and MeLU are not used in these datasets. Among all baseline models, TIGER and Mamba4Rec still achieve the best performance. Our model still outperforms all baselines for almost all the test cases except the recall scores. However, when taking both recall and precision to compute F1 scores, our model can still outperform all the baselines. Such improvement is mainly caused by our additional loss in our third module.

\end{document}